\documentclass[twocolumn]{elsart3p}
\usepackage[dvips]{graphicx}
\usepackage{bm}
\usepackage{amsmath}
\journal{Physica C}
\begin{document}

\begin{frontmatter}



\title{Field-angle dependence of the quasiparticle scattering inside a vortex core in unconventional superconductors}


\author[AA,BB,CC]{Y. Higashi\corauthref{cor1}}
\ead{higashiyoichi@ms.osakafu-u.ac.jp},\,
\author[CC,DD,EE]{Y. Nagai},\,
\author[CC,DD,EE]{M. Machida},\,
\author[BB,CC]{N. Hayashi}

\address[AA]{
Department of Mathematical Sciences, Osaka Prefecture University, 1-1 Gakuen-cho, Naka-ku, Sakai 599-8531, Japan
}

\address[BB]{
Nanoscience and Nanotechnology Research Center (N2RC), Osaka Prefecture University, 1-2 Gakuen-cho, Naka-ku, Sakai 599-8570, Japan
}

\address[CC]{
CREST(JST), 4-1-8 Honcho, Kawaguchi, Saitama 332-0012, Japan
}

\address[DD]{
CCSE, Japan Atomic Energy Agency, 6-9-3 Higashi-Ueno, Taito-ku, Tokyo 110-0015, Japan
}

\address[EE]{
TRIP(JST), 5 Sanban-cho, Chiyoda-ku, Tokyo 110-0075, Japan
}

\corauth[cor1]{
Address: N2RC, Osaka Prefecture University, C10 Bldg., 1-2 Gakuen-cho, Naka-ku, Sakai 599-8570, Japan
\\
Tel.: +81-72-254-9829 ; fax: +81-72-254-8203
\\
}

\newpage

\begin{abstract}
We theoretically investigate the quasiparticle scattering rate $\varGamma$ inside a vortex core
in the existence of non-magnetic impurities distributed randomly in a superconductor.
We show that the dependence of $\varGamma$ on the magnetic field direction
is sensitive to the sign of the pair potential.
The behavior of $\varGamma$ is quite different between an $s$-wave and a $d$-wave pair potential,
where these are assumed to have the same amplitude anisotropy, but a sign change only for  the $d$-wave one.
It is suggested that measurements of the microwave surface impedance with changing applied-field directions
would be used for the phase-sensitive identification of pairing symmetry.
\end{abstract}


\end{frontmatter}

\section{Introduction}
Much attention has been focused on unconventional superconductors \cite{sigrist}.
It is important to identify Cooper pair symmetry for getting a clue to pairing mechanism in each superconductor.
The Cooper pair is described by the superconducting pair potential, which is a complex number
with amplitude and phase.
Information on its phase is crucial for identifying Cooper pair symmetry
and for discriminating between unconventional superconductivity (e.g., $d$-wave pair)
and conventional one ($s$-wave pair).
However, most of experimental methods can detect only the amplitude of the pair potential,
whereas a few are phase sensitive.
Therefore, it is desirable to look for new methods capable of detecting the phase of the pair potential.

In this paper, we theoretically investigate the quasiparticle scattering rate $\varGamma$ inside a vortex core
and show that it can be used as a phase sensitive probe when changing the direction of applied magnetic field.
The field-angle dependences of the thermal conductivity and the specific heat
have been discussed so far \cite{field-angle,field-angle2}.
Instead, we will consider here the field-angle dependence of $\varGamma$ inside a vortex core.
The information on $\varGamma$ would be obtained experimentally
by measurements of the microwave surface impedance.

\section{Formulation}
We consider a single vortex core in a system in which non-magnetic impurities are distributed randomly.
The quasiparticle scattering rate $\varGamma$ is obtained
from the imaginary part of the impurity self energy ${\rm Im}\Sigma$.
We calculate $\Sigma$ on the basis of the quasiclassical theory of superconductivity \cite{serene}.
For the vortex bound states around a vortex core, analytic solutions of the quasiclassical Green's functions
are obtained by the so-called Kramer-Pesch approximation \cite{KP}.
$\Sigma$ is calculated from those Green's functions.
An $s$-wave scattering in the Born limit is assumed in this paper.
The pair potential is expressed as
$\Delta({\bm r}, {\bm k}_{\rm F})
={\tilde \Delta}({\bm r}) d({\bm k}_{\rm F})$,
where the Fermi wave number ${\bm k}_{\rm F}$ denotes the position on a Fermi surface
and the vortex center is situated at ${\bm r}=0$.
The function $d({\bm k}_{\rm F})$ represents the pairing symmetry.
The bulk amplitude of the pair potential  is defined by
$\Delta_0 = \bigl| {\tilde \Delta}( |{\bm r}| \to \infty) \bigr| $
away from a vortex core.
$\bm{v}_{\rm F}(\bm{k}_{\rm F})$ is the Fermi velocity at ${\bm k}_{\rm F}$.
Consider a scattering process of a quasiparticle from ${\bm k}_{\rm F}$ to ${\bm k}^\prime_{\rm F}$
on a Fermi surface.
$\varGamma$ is composed of an integration of the scattering process
with respect to the initial state ${\bm k}_{\rm F}$ and the final state ${\bm k}^\prime_{\rm F}$.
According to procedures described in Ref.\ \cite{nagai-kato},
the scattering rate $\varGamma(\varepsilon)$ for the vortex bound states with energy $\varepsilon$ is calculated as
\begin{eqnarray}
\frac{\varGamma(\varepsilon)}{\varGamma_{\rm n}}
&=&
\frac{\pi}{2}  
\Bigg\langle
\Bigg\langle
     \bigl(1-\mathop{\mathrm{sgn}}\nolimits[d(\bm{k}_{\rm F})d(\bm{k}^\prime_{\rm F})] \cos\Theta  \bigr) 
\nonumber \\
& & { } \qquad   \times \frac{1}{\vert\sin\Theta\vert}\frac{\vert \bm{v}_{\rm F \perp}(\bm{k}^\prime_{\rm F}) \vert}{\vert \bm{v}_{\rm F \perp}(\bm{k}_{\rm F}) \vert}\frac{\vert d(\bm{k}_{\rm F}) \vert}{\vert d(\bm{k}^\prime_{\rm F})\vert}
\nonumber \\
& & { } \qquad   \times e^{-u(s_0,\bm{k}_{\rm F})}e^{-u(s^\prime_0,\bm{k}^\prime_{\rm F})}
\Bigg\rangle_{FS^\prime}
\Bigg\rangle_{FS},
\nonumber \\
\label{eq:1}
\end{eqnarray}
where
%
%
\begin{eqnarray}
\Theta(\bm{k}_{\rm F}, \bm{k}^\prime_{\rm F})
\equiv
\theta_v(\bm{k}_{\rm F})-\theta_{v^\prime}(\bm{k}^\prime_{\rm F}),
\end{eqnarray}
%
\begin{eqnarray}
u(s,\bm{k}_F)=
\frac{2\vert d(\bm{k}_{\rm F})\vert\Delta_0}{\vert\bm{v}_{\rm F \perp}(\bm{k}_{\rm F})\vert}
\int_0^{\vert s \vert}ds^\prime\tanh\left(\frac{s^\prime}{\xi_0}\right),
\nonumber \\
\end{eqnarray}
%
\begin{eqnarray}
s_0(\bm{k}_{\rm F},\bm{k}^\prime_{\rm F},\varepsilon)
&=&
\frac{\varepsilon}{\sin\Theta}
\Biggl(
\frac{\vert \bm{v}_{\rm F \perp}(\bm{k}^\prime_{\rm F}) \vert}{2\Delta^2_0\vert d(\bm{k}^\prime_{\rm F}) \vert^2}
\nonumber \\
& & { } \qquad
-\frac{\vert \bm{v}_{\rm F \perp}(\bm{k}_{\rm F}) \vert}{2\Delta^2_0\vert d(\bm{k}_{\rm F}) \vert^2}\cos\Theta
\Biggr),
\nonumber \\
\end{eqnarray}
%
\begin{eqnarray}
s^\prime_0(\bm{k}_{\rm F},\bm{k}^\prime_{\rm F},\varepsilon)
&=&
s_0(\bm{k}_{\rm F},\bm{k}^\prime_{\rm F},\epsilon)\cos\Theta
\nonumber \\
& & { }
-\varepsilon\frac{\vert \bm{v}_{\rm F \perp}(\bm{k}_{\rm F}) \vert}{2\Delta^2_0\vert d(\bm{k}_{\rm F}) \vert^2}\sin\Theta.
\nonumber \\
\label{eq:5}
\end{eqnarray}
Here, $\varGamma_{\rm n}$ is the scattering rate in the normal state.
The brackets $\langle \cdots \rangle_{\mathop{\mathrm{FS}} }$ and
$\langle \cdots \rangle_{\mathop{\mathrm{FS}}^\prime}$ mean the Fermi-surface integrals
with respect to ${\bm k}_{\rm F}$ and ${\bm k}^\prime_{\rm F}$, respectively, like
\begin{equation}
\langle \cdots \rangle_{\mathop{\mathrm{FS}}^\prime}
\equiv \int  \frac{dS_{\rm F}(\bm{k}^\prime_{\rm F})}{\vert \bm{v}_{\rm F}(\bm{k}^\prime_{\rm F}) \vert} \cdots,
\end{equation}
with $dS_{\rm F}$ being an area element on a Fermi surface.

We have considered a single vortex in a coordinate system fixed to an applied magnetic field ${\bm B}$.
$\hat{\bm{a}}_{\rm M}$,
$\hat{\bm{b}}_{\rm M}$,
$\hat{\bm{c}}_{\rm M}$
are orthogonal unit vectors
and $\hat{\bm{c}}_{\rm M}$ is set parallel to ${\bm B}$ ($\hat{\bm{c}}_{\rm M} \parallel {\bm B}$).
$\bm{v}_{{\rm F}\bot}(\bm{k}_{\rm F})$ is the vector component
of $\bm{v}_{\rm F}(\bm{k}_{\rm F})$
projected onto the $a_{\rm M}$-$b_{\rm M}$ plane normal to ${\bm B}$.
Then, $\xi_0$ is defined by
$\xi_0=v_{{\rm F}\bot} / \pi\Delta_0$ with \cite{Nagai2007}
\begin{equation}
v_{{\rm F}\bot}
\equiv
\frac{
\int \frac{dS_{\rm F}(\bm{k}_{\rm F})}  {\vert \bm{v}_{\rm F}(\bm{k}_{\rm F}) \vert}
\vert \bm{v}_{{\rm F}\bot} (\bm{k}_{\rm F}) \vert
}{
\int \frac{dS_{\rm F}(\bm{k}_{\rm F})}  {\vert \bm{v}_{\rm F}(\bm{k}_{\rm F}) \vert}
}.
\end{equation}
Around the vortex, the amplitude of the pair potential was assumed to be
$|{\tilde \Delta}({\bm r})|=\Delta_0 \tanh(|{\bm r}|/\xi_0)$ in the $a_{\rm M}$-$b_{\rm M}$ plane.
$\theta_v(\bm{k}_{\rm F})$ is the angle of $\bm{v}_{{\rm F}\bot}(\bm{k}_{\rm F})$
measured from the $\hat{\bm{a}}_{\rm M}$ axis.

The expression of $\varGamma(\varepsilon)$ [Eqs.\ (\ref{eq:1})--(\ref{eq:5})]
is represented by two coordinate systems;
one is fixed to crystal axes and the other is fixed to the applied magnetic field.
That is, the wave number $\bm{k}_{\rm F}$ is defined in a coordinate system fixed to crystal axes,
while $\xi_0$ and the angle $\theta_v$ are defined in the  coordinate system fixed to the applied magnetic field.
Actually, there is a relation between those two coordinate systems.

We consider a coordinate system fixed to crystal axes characterized by
orthogonal unit vectors
$\hat{\bm{a}}$,
$\hat{\bm{b}}$,
$\hat{\bm{c}}$.
First, let us start with the situation where
the two coordinate systems coincide 
($\hat{\bm{a}}_{\rm M} \parallel \hat{\bm{a}}$,
$\hat{\bm{b}}_{\rm M} \parallel \hat{\bm{b}}$,
$\hat{\bm{c}}_{\rm M} \parallel \hat{\bm{c}}$),
and then
rotate the vector $\hat{\bm{c}}_{\rm M}$ (i.e., the direction of ${\bm B}$)
by the polar angle $\pi/2$ around the $\hat{\bm{b}}$ axis
(thus, $\hat{\bm{a}}_{\rm M}=-\hat{\bm{c}}$ and $\hat{\bm{c}}_{\rm M}=\hat{\bm{a}}$).
Next, we rotate $\hat{\bm{c}}_{\rm M}$ ($\parallel {\bm B}$)
by the azimuth angle $\alpha_{\rm M}$ around the $\hat{\bm{c}}$ axis
(see Fig.\ 1).
In this case, we have the relation
\begin{equation}
\left(
\begin{array}{c}
 \hat{\bm{a}}_{\rm M} \\
 \hat{\bm{b}}_{\rm M} \\
 \hat{\bm{c}}_{\rm M}
\end{array}
\right)
=
\left(
\begin{array}{ccc}
0 & 0 &-1  \\
-\sin\alpha_M &\cos\alpha_M & 0 \\
\cos\alpha_M &\sin\alpha_M & 0 
\end{array}
\right)
\left(
\begin{array}{c}
 \hat{\bm{a}} \\
 \hat{\bm{b}} \\
 \hat{\bm{c}}
\end{array}
\right).
\label{eq:8}
\end{equation}

In this paper,
we consider a single-band superconductor
with an isotropic spherical Fermi surface, where
$\bm{v}_{\rm F}(\bm{k}_{\rm F}) \parallel \bm{k}_{\rm F}$
and
$|\bm{v}_{\rm F}(\bm{k}_{\rm F})| \equiv v_{\rm F} ={\rm const.}$,
for clarity.
Using the above relation (\ref{eq:8}),
one can express $\bm{v}_{{\rm F}}(\bm{k}_{\rm F})$
in the coordinate system fixed to the magnetic field,
then evaluate $\bm{v}_{{\rm F}\bot}(\bm{k}_{\rm F})$,
and finally obtain the relations
\begin{eqnarray}
|\bm{v}_{{\rm F}\bot} (\bm{k}_{\rm F})|
&=&
v_{\rm F}
\sqrt{
\cos^2\theta_k + A^2 \sin^2\theta_k
},
\\
\cos\theta_v (\bm{k}_{\rm F})
&=&
\frac{ -v_{\rm F} }{ |\bm{v}_{{\rm F}\bot} (\bm{k}_{\rm F})| }
\cos\theta_k,
\\
\sin\theta_v(\bm{k}_{\rm F})
&=&
\frac{ v_{\rm F} }{ |\bm{v}_{{\rm F}\bot} (\bm{k}_{\rm F})| }
A \sin\theta_k,
\\
A
&=&
\sin\phi_k \cos\alpha_{\rm M} - \cos\phi_k \sin\alpha_{\rm M},
\nonumber \\
\label{eq:12}
\end{eqnarray}
where
$\bm{k}_{\rm F}
=
k_{\rm F} (
\cos\phi_k \sin\theta_k \hat{\bm{a}}
+
\sin\phi_k \sin\theta_k \hat{\bm{b}}
+
\cos\theta_k \hat{\bm{c}}
)$.
Thus, all quantities are represented in the coordinate system fixed to the crystal axes.
We have now reached to a position where we can calculate Eq.\ (\ref{eq:1})
to investigate the dependence of $\varGamma$ on the field-angle $\alpha_{\rm M}$.

\section{Results}
\begin{figure}[t]
\begin{center}
    \begin{tabular}{p{50mm}p{50mm}}
      \resizebox{50mm}{!}{\includegraphics{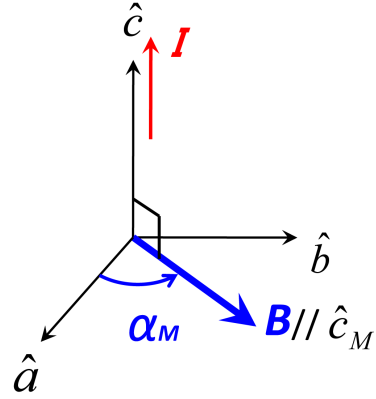}} 
    \end{tabular}
\end{center}
\caption{
A coordinate system fixed to crystal axes characterized by
orthogonal unit vectors
$\hat{\bm{a}}$,
$\hat{\bm{b}}$,
$\hat{\bm{c}}$.
The applied magnetic field ${\bm B}$ is rotated by the azimuth angle $\alpha_{\rm M}$ in the plane normal to the $\hat{\bm{c}}$ axis.
The $\hat{\bm{c}}_{\rm M}$ axis of the other coordinate system is taken parallel to ${\bm B}$.
A driving current ${\bm I}$ may be applied perpendicular to a plane in which ${\bm B}$ is rotated.}
\label{fig:1}
\end{figure}
\begin{figure}[t]
\begin{center}
    \begin{tabular}{p{80mm}p{80mm}}
      \resizebox{80mm}{!}{\includegraphics{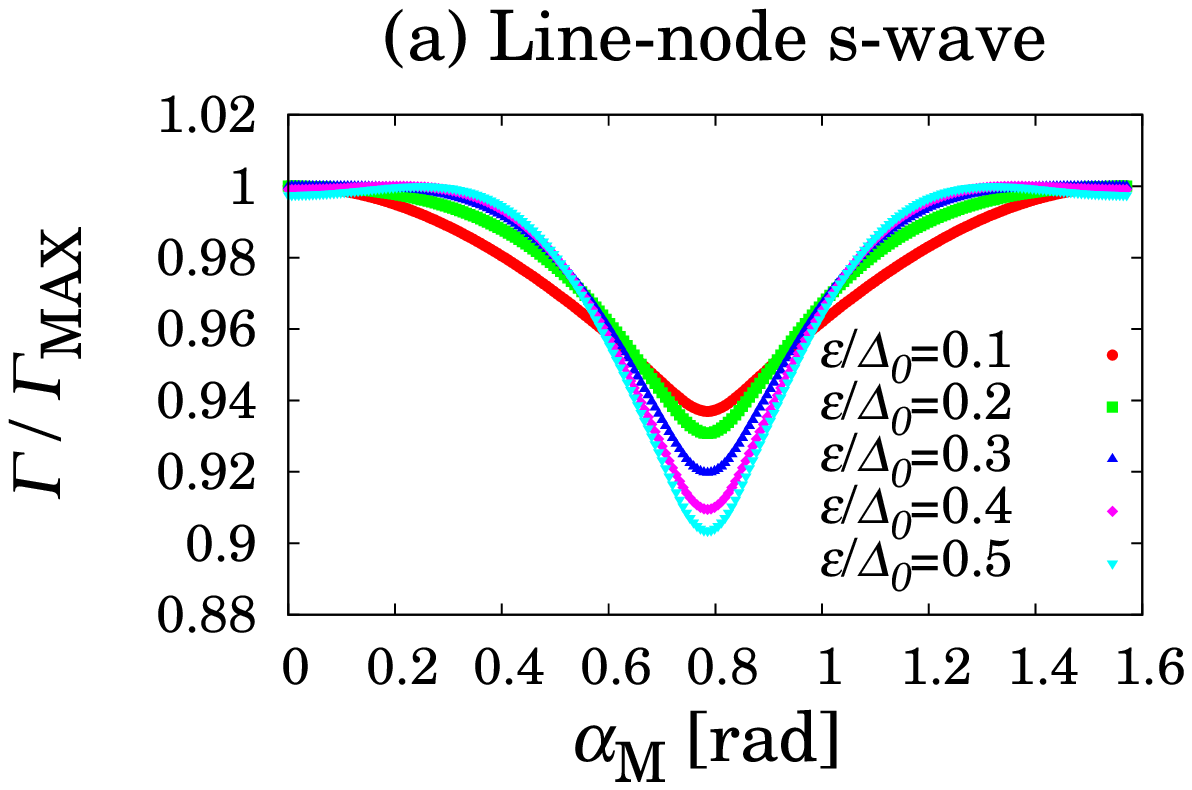}} \\
      \resizebox{80mm}{!}{\includegraphics{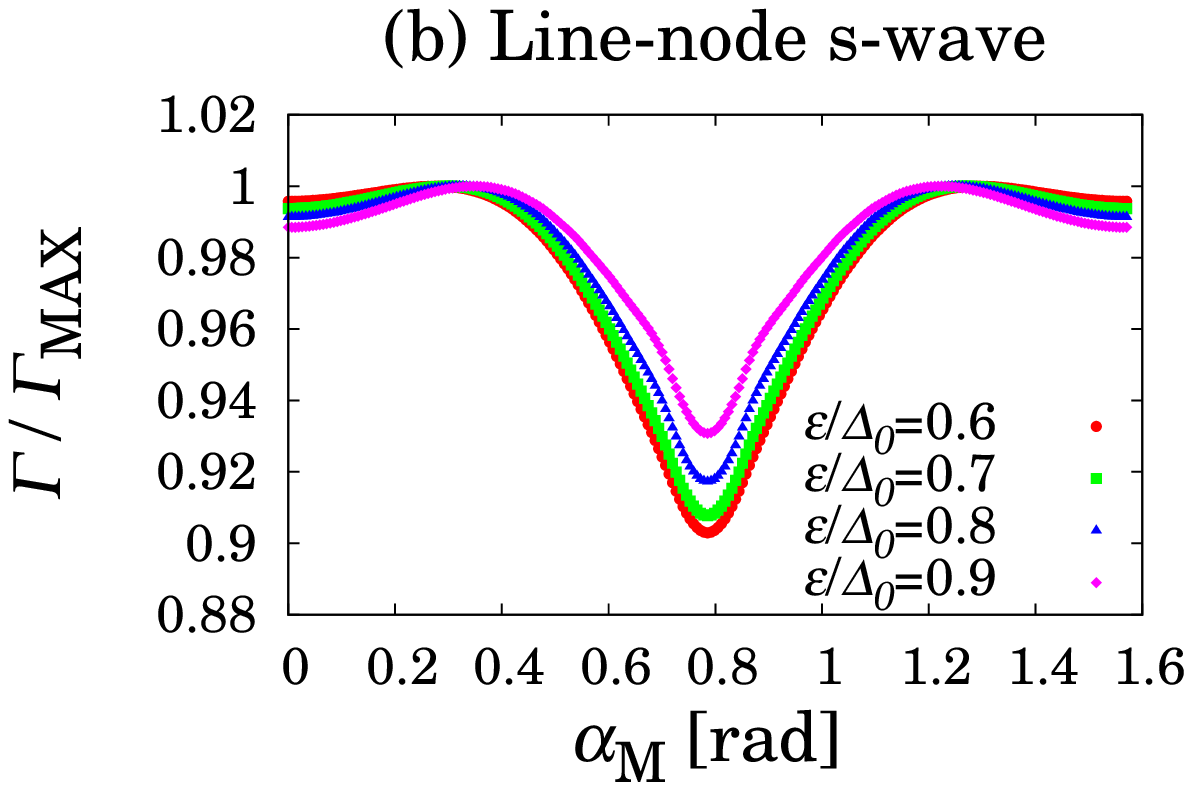}} 
    \end{tabular}
\end{center}
\caption{
Plot of the quasiparticle scattering rate $\varGamma$ vs.\ the applied-field angle $\alpha_{\rm M}$
in the case of the line-node $s$-wave pair.
Each curve is plotted for different quasiparticle energies $\varepsilon$.
The vertical axis is normalized by maximum value for each curve.}
\label{fig:2}
\end{figure}

\begin{figure}[t]
\begin{center}
    \begin{tabular}{p{80mm}p{80mm}}
      \resizebox{80mm}{!}{\includegraphics{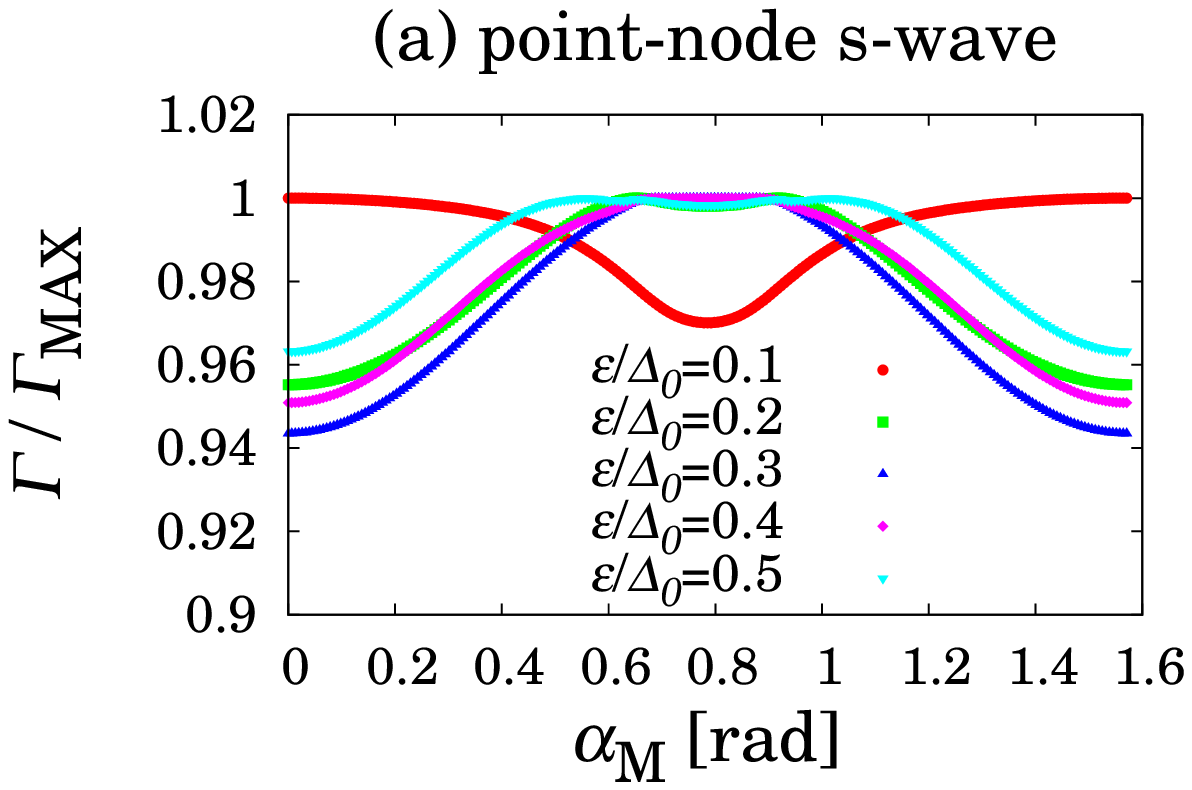}} \\
      \resizebox{80mm}{!}{\includegraphics{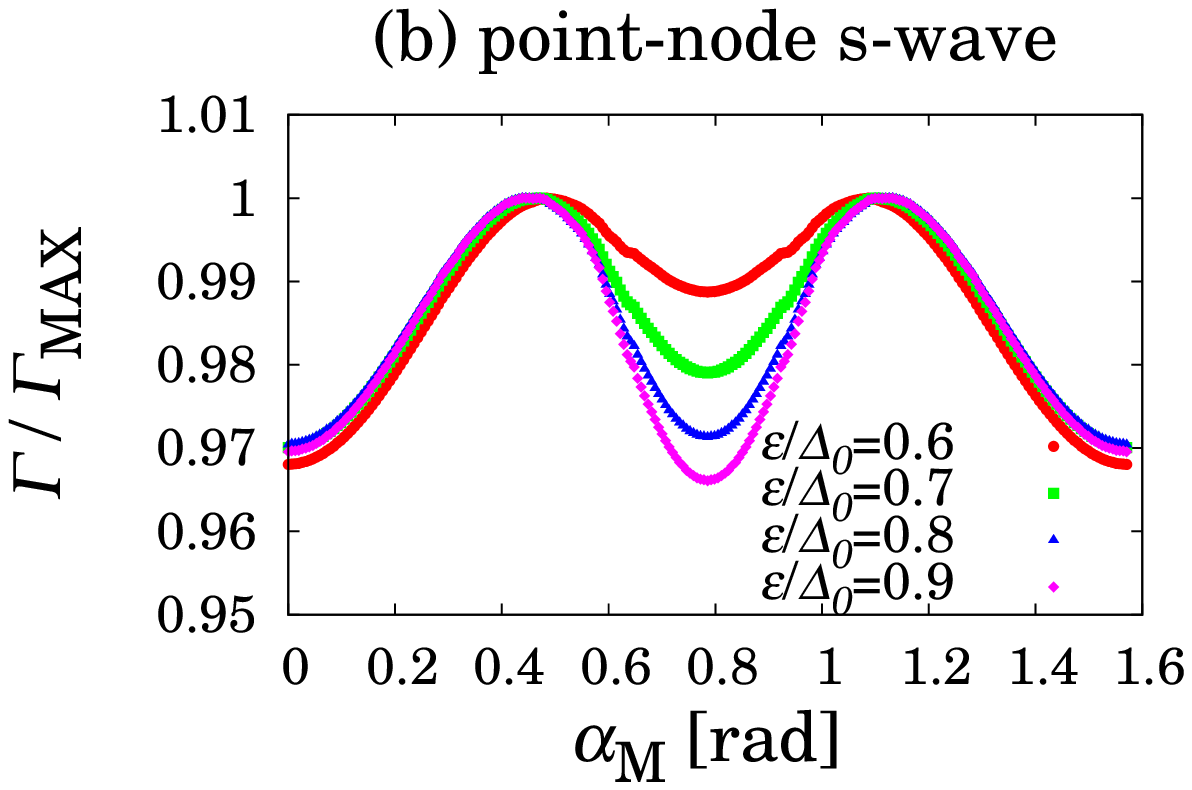}} 
    \end{tabular}
\end{center}
\caption{
Plot of the quasiparticle scattering rate $\varGamma$ vs.\ the applied-field angle $\alpha_{\rm M}$
in the case of the point-node $s$-wave pair.
Each curve is plotted for different quasiparticle energies $\varepsilon$.
The vertical axis is normalized by maximum value for each curve.}
\label{fig:3}
\end{figure}

\begin{figure}[t]
\begin{center}
    \begin{tabular}{p{80mm}p{80mm}}
      \resizebox{80mm}{!}{\includegraphics{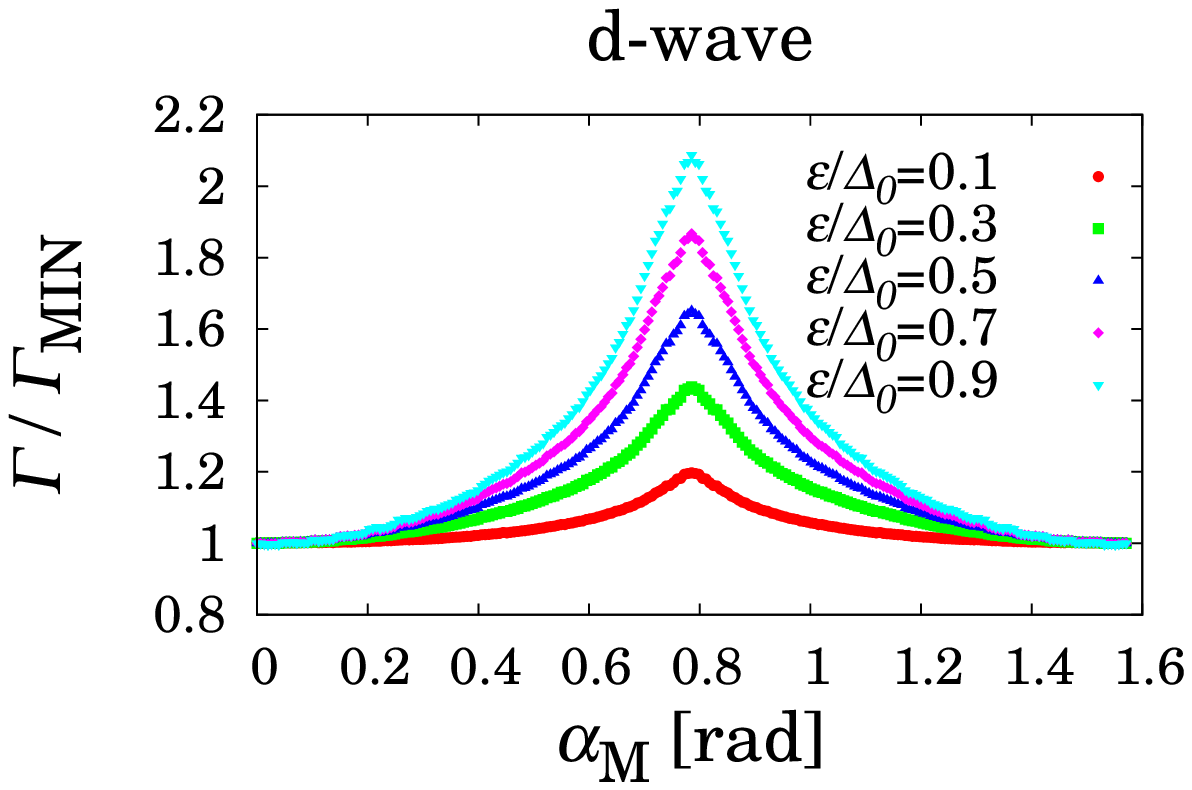}} 
    \end{tabular}
\end{center}
\caption{
Plot of the quasiparticle scattering rate $\varGamma$ vs.\ the applied-field angle $\alpha_{\rm M}$
in the case of the $d$-wave pair.
Each curve is plotted for different quasiparticle energies $\varepsilon$.
The vertical axis is normalized by minimum value for each curve.}
\label{fig:4}
\end{figure}

In this section, we show numerical results obtained from Eq.\ (\ref{eq:1}).
The following three types of Cooper pairing are considered on the isotropic spherical Fermi surface.
(i) Line-node $s$-wave:
$d(\bm{k}_{\rm F})=| \cos2\phi_k \sin^2\theta_k |$.
(ii) Point-node $s$-wave \cite{izawa}:
$d(\bm{k}_{\rm F})=(1+\cos4\phi_k \sin^4\theta_k )/2$.
(iii) $d$-wave:
$d(\bm{k}_{\rm F})=\cos2\phi_k \sin^2\theta_k $.
Note that all these pairing states have gap nodes ($|d(\bm{k}_{\rm F})|=0$)
in the $\phi_k=\pi/4$ directions on the Fermi surface.
On the other hand, the anti-node directions correspond to the $\phi_k=0$ ones.

In Figs.\ \ref{fig:2}--\ref{fig:4}, we show the scattering rate $\varGamma$
of the vortex bound states
as a function of the field-angle $\alpha_{\rm M}$ (see also Fig.\ \ref{fig:1}).
Each curve corresponds to different energies $\varepsilon$ of the quasiparticle.
In the case of the line-node $s$-wave pair (Fig.\ \ref{fig:2}),
$\varGamma$ exhibits its minimum around the gap-node direction $\alpha_{\rm M}=\pi/4$.
While similar behavior is seen also in the case of the point-node $s$-wave pair  (Fig.\ \ref{fig:3}) 
at the low energy $\varepsilon=0.1\Delta_0$, 
a difference appears at higher energies. 
The minimum appears also around the anti-node direction in the case of the point-node pair, which is in contrast to
the case of the line-node s-wave pair.
More prominent difference appears between
the line-node $s$-wave and the $d$-wave pair (compare Figs.\ \ref{fig:2} and \ref{fig:4}).
$\varGamma$ always exhibits its maximum at the gap-node direction in the case of the $d$-wave pair
and the ratio of the maximum value to the minimum one increases with increasing the energy $\varepsilon$,
which quite contrasts to the case of the line-node $s$-wave pair.

These two pair potentials, (i) and (iii) above,
have the same form in the amplitude $|d(\bm{k}_{\rm F})|$.
The only difference is that the pair potential exhibits a sign change for the $d$-wave pair (iii), and not for the $s$-wave one (i).
According to Eq.\ (\ref{eq:1}) of $\varGamma$,
the only part that reflects
the sign of the pair potential
is
the coherence factor \cite{nagai-kato},
$1-\mathop{\mathrm{sgn}}\nolimits[d(\bm{k}_{\rm F})d(\bm{k}^\prime_{\rm F})] \cos\Theta$.
Owing to this factor, arises the prominent difference in the field-angle dependence of $\varGamma$
between the $s$-wave and the $d$-wave pair state.

\section{Discussion}
As seen in the preceding section,
the field-angle dependence of the scattering rate $\varGamma$
of the vortex bound states
reflects the Cooper pairing symmetry.
This suggests that
its measurements can be considered as a new phase-sensitive probe.
$\varGamma(\varepsilon)$ of the vortex bound states
is related to the flux flow resistivity
$\rho_{\rm f}(T) \propto \varGamma(\varepsilon=k_{\rm B}T)$
in moderately clean systems \cite{kato}
even when the superconducting gap has nodes on a Fermi surface \cite{kopnin}.
The field-angle dependence of $\rho_{\rm f}$ may be measured by rotating an applied magnetic field
within a plane normal to a driving current (Fig.\ \ref{fig:1}) \cite{yasuzuka}.
The microwave surface impedance measurement is a promising experimental method to estimate $\rho_{\rm f}$
avoiding the vortex pinning effect at low temperatures \cite{micro,micro2} if it is possible to rotate
an applied magnetic field there.

\section{Conclusion}
We have investigated the quasiparticle scattering rate $\varGamma$ inside a vortex core
under the existence of non-magnetic impurities.
The expression of $\varGamma$ is derived in a form where the applied magnetic field can be rotated in a plane.
We find that the behavior of $\varGamma$ as a function of the applied-field direction reflects
the sign of the Cooper pairing.
The difference in  $\varGamma$ between the $s$-wave and the $d$-wave pair  is especially prominent
despite the same gap amplitude anisotropy they have.
That difference arises because of the coherence factor which depends on the sign of the pair potential.
Therefore, the field-angle dependence of $\varGamma$ is sensitive to the phase of Cooper pair
in contrast to the field-angle dependence of the specific heat, which is known to be phase insensitive.
Elucidation of more detailed mechanism of the field-angle dependences in $\varGamma$ is left for future studies.
$\varGamma$ is related to the flux flow resistivity, the field-angle dependence of which
would be experimentally explored by microwave surface impedance measurements
if the direction of an applied magnetic field can be changed there.

\section*{Acknowledgments}
We would like to thank S. Yasuzuka, Y. Kato, K. Izawa, and M. Kato for helpful discussions.
%

\newpage


\end{document}